\documentclass[twocolumn,prd,aps,tightenlines,floats,floatfix,preprintnumbers,
nofootinbib,eqsecnum]{revtex4}

\def\sp{\kern +3pt}
\def\sm{\kern -3pt}
\def\spQ{\kern +6pt}

\def\bea{\begin{eqnarray}}
\def\eea{\end{eqnarray}}

\def\sfrac#1#2{{\textstyle \frac{#1}{#2}}}

\def\be{\begin{equation}}
\def\ee{\end{equation}}
\def\ba{\begin{eqnarray}}
\def\ea{\end{eqnarray}}

\usepackage{graphics}
\usepackage{graphicx}
\usepackage{epsf}
\usepackage{amsmath}
\usepackage{amssymb}
\usepackage{bm}

\usepackage{color}

\usepackage[switch]{lineno}
\usepackage{blindtext}
\usepackage{lipsum}

\usepackage{lineno}

\begin{document}

\phantom{0}
\vspace{-0.2in}
\hspace{5.5in}

\preprint{\bf LFTC-19-6/44}

\vspace{-1in}

\title
{\bf Octet baryon electromagnetic form factor double ratios \\
$(G_E^\ast/G_M^\ast)/(G_E/G_M)$  in a nuclear medium}
\author{G.~Ramalho, J.~P.~B.~C.~de Melo, and K.~Tsushima}
\vspace{-0.1in}

\affiliation{Laborat\'orio de 
F\'{i}sica Te\'orica e Computacional -- LFTC,
Universidade Cruzeiro do Sul/Universidade Cidade de S\~ao Paulo, \\
01506-000, S\~ao Paulo, SP, Brazil}

\vspace{0.2in}
\date{\today}

\phantom{0}

\begin{abstract}
The ratios of the baryon electric ($G_E$)
to magnetic ($G_M$) form factors, $G_E/G_M$, 
can provide us with important information 
on the structure of baryons in vacuum as demonstrated in recent studies 
for the proton and neutron systems.
It has been argued that the corresponding in-medium ratios, $G_E^\ast/G_M^\ast$, 
can also provide useful information on the electromagnetic structure 
of baryons in a nuclear medium. 
Although the ratios may not be measured directly in experiments 
except for the proton at present, 
the double ratios $(G_E^\ast/G_M^\ast)/(G_E/G_M)$ 
can be directly related to the polarization-transfer 
measurements of bound baryons. 
In the present work we estimate, for the first time, the 
double ratios $(G_E^\ast/G_M^\ast)/(G_E/G_M)$  
for all the members of octet baryons in symmetric nuclear 
matter in the range $Q^2=0$--3 GeV$^2$,
where $Q^2=- q^2$ with $q$ being the four-momentum transfer.
\end{abstract}

\vspace*{0.9in}  
\maketitle

\section{Introduction}

The issue of whether or not hadrons change their properties in a nuclear medium, 
has been one of the long-standing problems in nuclear 
physics~\cite{Brown91,Saito07,Brooks11}.
Quantum chromodynamics (QCD) is established as the theory of strong interaction, 
and quarks and gluons are the degrees of freedom in the QCD Lagrangian. 
It is natural that in the strong mean fields which 
pervade nuclear matter, the motion of the quarks and gluons 
inside hadrons should be modified. 
Such changes are what is meant by
the nuclear modification of hadron properties, and studying the effects 
is clearly central to the understanding of hadron properties 
in a nuclear medium within QCD.

In the last few decades new experimental methods have been proposed to 
probe the electromagnetic structure of baryons in a nuclear medium.
 The measurements made at JLab~\cite{NSTAR} and 
Mainz Microton (MAMI) for polarization-transfer 
double ratio for a bound proton, which can be directly connected  
with the electromagnetic form factor (EMFF) double ratio  
$(G_E^\ast/G_M^\ast)/(G_E/G_M)$~\cite{Dieterich01,Strauch04,Paolone10,Strauch10a,Strauch18},
showed that the ratio is not unity, and imply the medium modifications of 
the bound proton electromagnetic form factors.

The $G_E^\ast/G_M^\ast$ ratio is measured 
based on the relation between the polarization-transfer
coefficients in a quasielastic reaction 
where a proton is knocked out from a nucleus (quasifree proton).
Although the final proton is in vacuum,
one can still assume that the polarization-transfer
coefficients carry the information  
of the bound proton, since the photon coupling 
with the proton occurs in a nuclear environment, 
and the polarization is conserved.
This interpretation has been used by many different groups 
in the studies of the electromagnetic structure of the
nucleon in nuclear medium~\cite{Frank96,Octet2,Cloet09,Kim16,deAraujo18,Lu99a,PAC35}.
Although there are still debates in the interpretation 
of the measured polarization-transfer coefficients 
as representing the proton electromagnetic properties in a nuclear medium, 
since the knocked out proton is detected as a quasifree particle in vacuum,
this is presently considered to be the best and cleanest method 
to extract the effect of the nuclear medium for 
the bound proton electromagnetic form factors.

The results from JLab and MAMI motivated 
further theoretical studies  
of the double ratios, not only for the proton but also the neutron, using different
frameworks~\cite{Frank96,Octet2,Cloet09,Kim16,deAraujo18,Vega18,Lu99a,Yakhshiev03,Smith04,Horikawa05,Cloet16}.
Although model dependent, these studies predict a surprising fact that, while 
the double ratio for the proton is quenched, 
that for the neutron is enhanced in a nuclear medium~\cite{Octet2,Cloet09,deAraujo18}. 
Thus, one can suspect that the in-medium modifications of the electromagnetic form factors 
may be different for the charged and charge neutral particles.
One of the important and interesting motivations of this study is, 
to explore systematically the medium modifications of the octet baryon 
electromagnetic form factors focusing on the differences between 
the charged and charge neutral baryons.

In high-energy heavy-ion collisions and in the core of a neutron star and 
a compact star, the formation of baryon-nucleus bound states are less expected  
due to the disappearance of nuclei into nuclear, hadron, 
or quark matter. However, for matter produced in heavy-ion collisions and 
expected to exist in the core of neutron and compact stars, 
the medium effects that are similar to the system of 
the baryon-nucleus bound states may be produced.
Thus, as a first step, we consider the situation that the octet baryons 
are immersed in symmetric nuclear matter and study the medium effects 
on the electromagnetic form factors of them. 
Such a study may provide us with very important information 
for understanding better heavy-ion collisions, 
neutron stars and compact stars.

In the present work we estimate the EMFF double ratios 
of octet baryons in symmetric nuclear matter. 
In addition to the nucleons, we predict   
the double ratios of the charged baryons  
$\Sigma^+$, $\Sigma^-$, and $\Xi^-$,
as well as the charge neutral baryons $\Lambda$, $\Sigma^0$, and $\Xi^0$.
We consider, in particular, nuclear matter densities $\rho=0$ and 
$0.5 \rho_0$, and $\rho_0$,
with $\rho_0 = 0.15$ fm$^{-3}$ being the normal nuclear matter density.
The present estimates are based on the model results in Ref.~\cite{Octet2}.
In this work, however, we estimate directly the double ratios    
$(G_E^\ast/G_M^\ast)/(G_E/G_M)$ of octet baryons, 
since these quantities can possibly be compared with the polarization-transfer 
double ratio measurements, when the experimental technique advances 
in the future.

Our estimates are based on the combination of
the covariant spectator quark model and the quark-meson coupling (QMC) model, 
extended in symmetric nuclear matter~\cite{Octet2}.
The free parameters of the covariant spectator quark model 
in vacuum are calibrated using lattice QCD EMFF data 
for the octet baryons avoiding the contamination  
from the meson cloud effects. 
The meson cloud contributions
for the EMFFs are estimated based on an 
$SU(6)$ symmetry for the meson-baryon couplings.
Since the meson cloud is expected to be dominated 
by the lightest meson, the pion, we reduce 
the calculations to the pion cloud contributions.
The parameters associated with the pion cloud are fitted 
to the available physical data for 
the octet baryons including the nucleon 
EMFFs, octet baryon magnetic moments, 
and the electric and magnetic radii.
The extension for symmetric nuclear matter was already 
made in Ref.~\cite{Octet2} taking into account the medium modifications  
of relevant hadron masses and pion-baryon coupling constants 
combined with the QMC model~\cite{Lu99a,Saito07}.
The study of the properties of the nuclear matter 
is performed in the nuclear matter rest frame.
Then, the nuclear matter densities considered 
in the present study refer to the nuclear matter rest frame.
Note that, although the in-medium quantities used in this study are calculated 
in the rest frame of nuclear matter in the QMC model, 
they are all Lorentz-scalar quantities.
The parameters of the QMC model relevant for describing  
nuclear matter, the coupling constants between the light quarks 
and the Lorentz-scalar as well as the Lorentz-vector mean fields, 
are fixed so as to reproduce the empirically extracted nuclear matter 
saturation properties, namely, the binding energy  
and the asymmetry energy coefficient at the saturation density point.
 
This article is organized as follows.
In Sec.~\ref{secFF}, we introduce nomenclature 
associated with the octet baryon electromagnetic 
form factors in vacuum and in medium.
In Sec.~\ref{secModel}, we discuss briefly 
the properties of the theoretical model used,  
and explain the model extension for symmetric nuclear matter.
The results for the octet baryon EMFF double ratios 
are presented and analyzed in Sec.~\ref{secResults}.
Finally, we give outlook and conclusions in Sec.~\ref{secConclusions}.

\section{Electromagnetic form factors in vacuum and in medium}
\label{secFF}

The electromagnetic current for the interaction between   
an octet baryon $B$ and a virtual photon with momentum $q$
in vacuum, can be written as
\ba
J_B^\mu = F_{1B} (Q^2) \gamma^\mu + 
F_{2B} (Q^2) \frac{i \sigma^{\mu \nu} q_\nu}{2 M_B},
\label{eqJB1}
\ea
where $M_B$ is the baryon mass,
$Q^2=-q^2$ is the photon virtuality,
and,  $F_{1B}$ and $F_{2B}$ are 
the Dirac and Pauli form factors, respectively.
In Eq.~(\ref{eqJB1}) we represent the current 
in units $e= \sqrt{4\pi \alpha}$,
where $\alpha \simeq 1/137$ is the electromagnetic 
fine structure constant.
For simplicity, we omit the projection with the initial $u_B$ 
and final $\bar{u}_B$ Dirac spinors. 

At $Q^2=0$ the baryon form factors are 
normalized as $F_{1B}(0)=e_B$ and  $F_{2B}(0)=\kappa_B$,
where $e_B$ is the baryon charge in units of $e$, and 
$\kappa_B$ is the baryon anomalous magnetic moment in natural units 
$\frac{e}{2 M_B}$.

An alternative representation of EMFFs for a spin-1/2
particle is the Sachs representation:
the electric $G_{EB}$ and magnetic $G_{MB}$ form factors.
These form factors are defined by~\cite{Octet2}
\ba
G_{EB} (Q^2) & \equiv & F_{1B} (Q^2)- \frac{Q^2}{4 M_B^2} F_{2B} (Q^2), 
\label{eqGEB} \\
G_{MB} (Q^2) & \equiv & \left[F_{1B} (Q^2) + F_{2B} (Q^2) \right] 
\frac{M_N}{M_B}.
\label{eqGMB}
\ea
Note that the expression for $G_{MB}$ includes 
the factor $\frac{M_N}{M_B}$, where $M_N$ is the free nucleon 
mass, in order to convert to the nucleon natural units ($\frac{e}{2 M_N}$).

If we omit the conversion factor $\frac{M_N}{M_B}$, 
we have the usual relation,  $G_{MB} = F_{1B} + F_{2B}$
in the baryon $B$ natural units ($\frac{e}{2 M_B}$), 
and in the limit $Q^2=0$ we obtain the 
baryon magnetic moment $\mu_B = G_{MB} (0) \frac{e}{2 M_B}$.
To summarize, the inclusion of the extra factor $\frac{M_N}{M_B}$ in Eq.~(\ref{eqGMB}) 
is to convert $G_{MB}$ to the nucleon natural units,  
allowing a simpler comparison among the magnetic form factors 
of different baryons.

We consider now the situation that baryons are immersed in 
symmetric nuclear matter.
By assuming a spin-1/2 baryon $B$ in symmetric nuclear matter  
as a quasiparticle which has an effective mass (Lorentz-scalar)
and an effective vector potential or four-momentum (Lorentz-vector) 
due to the medium effects satisfying the quasiparticle Dirac equation, 
we can also define the bound baryon 
$B$ EMFFs in a nuclear medium~\cite{Walecka} 
as $F_{1B}^\ast$ and $F_{2B}^\ast$, respectively, corresponding 
to those in vacuum without the asterisk *.
The Sachs form factors $G_{EB}^\ast$ and $G_{MB}^\ast$ 
in medium can also be obtained using the corresponding relations  
of Eqs.~(\ref{eqGEB}) and (\ref{eqGMB}), 
with the baryon $B$ effective mass $M_B^\ast$.
The effective mass $M_B^\ast$ appears also in the 
baryon $B$ spinor in medium. 

For convenience we use also 
$e_B^\ast = F_{1B}^\ast (0)$ and 
$\kappa_B^\ast = F_{2B}^\ast (0)$ 
to represent the charge and anomalous magnetic moments in medium, 
where $e_B^\ast = e_B$. 

Note in particular that, according to our convention, 
the in-medium magnetic form factor takes the form  
\ba
G_{MB}^\ast (Q^2) & \equiv & \left[F_{1B}^\ast (Q^2) + F_{2B}^\ast (Q^2) \right] 
\frac{M_N}{M_B^\ast}.
\ea
In this way, we represent the octet baryon in-medium magnetic 
form factors in natural units of the nucleon in vacuum,  
where $M_N$ is the free nucleon mass. 
For the magnetic moment of an octet baryon $B$ in medium 
we use $\mu_B^\ast = G_{MB}^\ast (0) \frac{e}{2M_N}$. 

Since the effective baryon mass   
is expected to be smaller than that in vacuum 
($M_B^\ast  < M_B$), $G_{MB}^\ast (Q^2)$ is expected 
to be enhanced in medium relative to that in vacuum.
In particular, one has $|\mu_B^\ast| > |\mu_B|$.

As for the electric form factor, since there is no  
additional factor, the charges of the baryons are not modified 
by the medium effects (charge conservation).

\section{Description of the model}
\label{secModel}

We explain in this section briefly the properties of the model used in the present work.
We start with the description of the model in vacuum, and, 
in the end, discuss the extension to the nuclear medium.

\begin{figure}[t]
\includegraphics[width=2.5in]{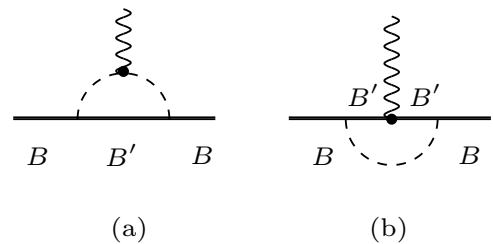}
\caption{\footnotesize
Electromagnetic interaction of a baryon $B$ 
with a photon within the one-pion loop level 
through the intermediate baryon states $B^\prime$.
}
\label{figPionCloud}
\end{figure}

The electromagnetic interaction of a baryon $B$ with a photon 
can be decomposed into the photon interaction with the valence quarks 
and the photon interaction 
with the (bare baryon)-meson system.
Since the pion is the lightest meson, 
we consider only the pion cloud effects as a first, reasonable approximation.
Then, we decompose the electromagnetic current into
\ba
J_B^\mu = Z_B \left[  
J_{0B}^\mu + J_\pi^\mu + J_{\gamma B}^\mu
\right],
\label{eqJtot}
\ea
where $J_{0B}^\mu$ stands for the electromagnetic 
interaction with the baryon core (valence quarks without pion cloud)
and the remaining terms represent the interaction 
with the intermediate pion-baryon ($\pi B$) states,
displayed in Fig.~\ref{figPionCloud}.
In particular, $J_\pi^\mu$ represents the direct interaction 
with the pion [Fig.\ref{figPionCloud}(a)] 
and the $J_{\gamma B}^\mu$ the interaction 
with the baryon while one pion is in the air [Fig.~\ref{figPionCloud}(b)].
$Z_B$ is a normalization constant  
associated with the baryon $B$ bare wave function.
For more details, see Refs.~\cite{Octet1,Octet2,OctetMM,LambdaSigma0}.

\subsection{Valence quark contribution}

The valence quark contributions for the octet baryon 
electromagnetic form factors are determined 
by the current $J_{0B}^\mu$ based on the 
covariant spectator quark model in Refs.~\cite{Octet1,Octet2}.

In the covariant spectator quark model 
the structure of the baryon wave functions 
is based on an $SU(3)$ symmetry, 
which is broken by the quark masses, namely, by the heavier strange quark mass.
In the relativistic impulse approximation  
we can separate the degrees of freedom of the interaction-acting  
quark from the spectator quark-pair (diquark) 
and integrate into the internal degrees of freedom of the quark-pair 
to obtain an effective 
radial wave function~\cite{Spectator-Review,Nucleon,Omega,Nucleon2,Nucleon3,OctetDecuplet1}.
Then, the $SU(3)$ symmetry breaking in the model is included at the level 
of the octet baryon radial wave functions approximated by an $S$-wave state 
of the quark-diquark system~\cite{Octet1,Octet2}.
The octet baryon radial wave functions are characterized by 
the Hulthen form associated with the two momentum-range scales
and depend also on the baryon mass ($M_B$)~\cite{Nucleon}.
We consider a common long-range scale for all octet baryons  
and a shorter-range scale associated with the number 
of strange quarks, namely, zero, one, or two~\cite{Octet1,Octet-Axial}.
Since we use different momentum-range scales 
for the different octet baryon families,
we break the $SU(3)$ symmetry at the level of the radial wave functions. 
Overall, one has four momentum-range scales 
for the octet baryon radial wave functions.
The  $SU(3)$ symmetry breaking effect has a strong impact 
on the baryon systems with two strange quarks. 

The electromagnetic structure of the valence constituent quarks 
is described by effective quark form factors,  
which can be parametrized in terms of 
the vector meson dominance (VMD) mechanism.

For the flavor $SU(2)$ sector, we represent 
the quark current based on the isoscalar (isovector)
decomposition into an $\omega$-meson ($\rho$-meson) associated term 
which parametrizes the long-range structure,  
and a term associated with a heavy effective meson 
(mass $M_h$) to simulate the short-range structure.
In numerical calculation we  
approximate $m_\omega \simeq m_\rho$,  
and fix $M_h$ as $M_h = 2 M_N$  (twice the nucleon mass).

For the strange quark, we replace the $\rho$-term 
($\pi \pi$ channel) by a term associated 
with the $\phi$ meson ($K \bar K$ channel).
One obtains in the end a quark current with 
five adjustable parameters, combined 
with three bare-quark anomalous magnetic moments, 
$\kappa_u$, $\kappa_d$ and $\kappa_s$.
The five parameters were determined     
in the study of the nucleon physical  
EMFFs~\cite{Nucleon}, and in the study of 
decuplet baryon EMFFs in lattice QCD~\cite{Omega}.
The quark bare-anomalous magnetic moments 
are readjusted in the study of 
octet baryon EMFFs in Ref.~\cite{Octet2}.

The covariant spectator quark model can be extended 
to the lattice QCD regime.
In this extension, we generalize the quark currents and the 
radial wave functions to the lattice QCD conditions,
in general, characterized by the value of the pion mass. 
The extension to lattice QCD regime assumes 
that the effect of the meson cloud is suppressed 
for simulations with large pion masses
and that only the effect of the valence quark is relevant.

In the extension to the lattice QCD regime,
the radial wave function 
is determined using the baryon mass $M_B$ 
associated with the lattice QCD regime 
and the quark currents are determined 
using a VMD parametrization where 
the physical masses ($M_N$, $m_\rho$, $m_\phi$ and $M_h= 2M_N$) 
are replaced by the corresponding masses 
in the lattice QCD regime~\cite{Lattice,LatticeD}. 
The parameters of the wave function and 
the parameters of the quark current
are the same in the physical and in the lattice QCD regimes. 
The extension to lattice QCD regime 
was proved to be successful for the nucleon 
and $\Delta(1232)$ systems~\cite{Lattice,LatticeD}. 
The method has been used in several studies 
of the octet and decuplet baryons in 
order to estimate the valence quark contributions 
for the EMFFs~\cite{Octet1,Octet2,Omega,Omega2}.

An important characteristic of the 
covariant spectator quark model is that
it breaks $SU(2)$ symmetry in the $u$ and $d$ 
light-quark sector at the level of quark current.
In the present model the strength of the isoscalar current 
is larger than that of the isovector compared 
with the other models~\cite{Chung91,Smith04,Thomas84,Miller02}.
This isovector-isoscalar asymmetry has 
an impact on the results of the neutron electric form factor 
and is responsible for the dominance of the   
valence quark component in the neutron electric 
form factor~\cite{Nucleon,Octet2}.

\subsection{Pion cloud contribution}

The pion cloud contributions for the octet baryon EMFFs 
are estimated using an $SU(3)$ symmetry-based model 
for the pion-baryon interaction parametrized 
by the constant $\alpha \equiv D/(D+F) = 0.6$ [$SU(6)$ symmetry-limit value].
We use the framework of Ref.~\cite{OctetMM},  
where all the couplings ($\pi B$, $eB$ and $\kappa B$)
are determined by the $SU(6)$ symmetry model, 
apart the six independent functions of $Q^2$~\cite{Octet2}.
Those functions are parametrized in terms 
of five independent coefficients, and two cutoffs 
which regulate the falloff of the pion cloud contributions~\cite{Octet1,Octet2}.
In the parametrizations we include 
a dependence on the pion mass in order to reproduce 
the leading order chiral behavior of 
the nucleon electromagnetic isovector square charge radii,  
corresponding to the Dirac and Pauli EMFFs~\cite{Octet2}.

\subsection{Calibration of the model in vacuum}

The parameters of the present model in vacuum  
are calibrated using lattice QCD EMFF data
for the baryons $p$, $n$, $\Sigma^+$, 
$\Sigma^0$, $\Sigma^-$ and $\Xi^-$~\cite{Lin09}.
Only for $\Lambda$ and $\Xi^0$ 
are there no representative lattice QCD
data\footnote{The lattice QCD simulations are more difficult to perform 
for charge neutral baryons, and thus have poorer statistics.
For these reasons the neutral baryons are in general 
excluded from the simulations.
In the present case there are results for $n$ and $\Xi^0$, 
although errors are large~\cite{Lin09}.}.
In addition to the lattice QCD data, 
we use also physical data to constrain the model in vacuum, 
namely, EMFF data for $p$ and $n$, the magnetic moments of   
octet baryons except for $\Sigma^0$, 
which is unknown, and the electric square charge radii of $p$, $n$ 
and $\Sigma^-$ as well as the magnetic square radii of $p$ and $n$.
The physical data are used to fix the parameters 
associated with the pion cloud, while the lattice data are used 
to fix the parameters associated with the valence quarks.
The normalization constant $Z_B$ is determined 
by the combination of the bare and pion cloud components.
Details of the model calibration can be found in Ref.~\cite{Octet2}.

\subsection{Extension to the nuclear medium}

As discussed already, for the extension of the model in vacuum for 
the nuclear medium, we assume that the proprieties of the octet baryons 
are modified in medium, and they are described by the  
modified Dirac equations for quasiparticles  
with the modified masses ($M_B^\ast$),  
and the modified pion-baryon coupling constants ($g_{\pi B B'}^\ast$).
The medium modifications of baryon masses (Lorentz-scalar)  
and coupling constants $g_{\pi B B'}^\ast$ are calculated based on the QMC 
model~\cite{Saito07} in the rest frame of nuclear medium. 
Since the effective baryon masses are Lorentz-scalar, 
and $g_{\pi B B'}^\ast$ are estimated using the corresponding 
effective baryon and pion masses, the Lorentz covariance is preserved.

The medium modifications of the valence quark contributions 
for the EMFFs are estimated as follows:
the effects of masses $M_B^\ast$ are included in  
the octet baryon radial wave functions $\Psi_B$  
and in the quark electromagnetic (EM) currents.
In the extension of the quark EM currents to the nuclear medium, 
we take the advantage of the VMD parametrization in vacuum  
with the free masses $m_\rho (\simeq m_\omega)$, $m_\phi$ and $M_h = 2 M_N$, 
and in medium we replace these masses by the respective values in medium,   
$m_\rho^\ast (\simeq m_\omega^\ast)$, $m_\phi^\ast$ and $M_h^\ast = 2 M_N^\ast$.

Concerning the medium modification of the pion cloud 
contribution, it is performed taking into account 
the medium modification of the pion-baryon coupling constants $g^*_{\pi B B'}$.
For estimating the in-medium pion-baryon coupling constants $g^*_{\pi B B'}$, 
we use the Goldberger-Treiman relation and an estimate of the in-medium pion decay 
constant $f_\pi^\ast$~\cite{Octet2}.  

The numerical values for the in-medium baryon ($M_B^\ast$) 
and meson masses, and coupling constants ($g_{\pi B B'}^\ast$) 
used in the present work, can be found 
in Tables~3 and 4 in Ref.~\cite{Octet2}, respectively.

\subsection*{Properties of the model in medium}

Before presenting the results for 
the $G_E/G_M$ double ratios of octet baryons, 
it is important to review briefly the properties of the model 
derived and calibrated in Ref.~\cite{Octet2}.
This will help better to understand 
the results of double ratios presented in this article.

From the calibration of the model in vacuum~\cite{Octet2}, 
it is clear that the fit to the lattice data provides 
a good description for the $N$ and $\Sigma$ systems 
but not so good for the $\Xi$ systems.
This may be the consequence of the $SU(3)$ symmetry 
breaking, which has a stronger impact on the systems 
with two strange quarks.

It was also concluded that the calibration provided only a crude
estimate of the quark core effect for the charge neutral baryons, 
since these systems are not well constrained by lattice QCD data,
except for the neutron and the $\Xi^0$. 
For the neutron, fortunately, there are physical data to be used. 
Nevertheless, the separation between the
quark core and pion cloud contributions should be taken with care,  
particularly for $G_{EB}$, whose magnitudes  
are in general small for charge neutral baryons, 
where the small pion cloud effects and residual quark core effects 
cannot be separated with a good precision.
The estimates for the charge neutral baryons can  
in principle be improved with the explicit inclusion 
of the kaon cloud as well.
Although the effect of the kaon cloud is expected 
to be smaller than that of the pion due to the large kaon mass, 
it may still be relevant in the cases 
when both the bare quark core and the 
pion cloud contributions are small. 

Based on the $SU(3)$ scheme applied in the present model,  
we conclude, in general, that the valence quark contribution 
is more than 80\% of the total contribution for each octet baryon
EMFF in vacuum, as well as in medium~\cite{Octet2}.
The exceptions are the charge neutral baryons 
which deserve specific discussions.

In general we notice that the ratio
$G_{E}^*/G_{E}$ decreases from unity as $Q^2$ increases  
except for the charge neutral baryons, which
means that the magnitudes of the electric form factors decrease
in symmetric nuclear matter faster than those 
in vacuum in the low-$Q^2$ region.
[See Figs.~4, 7, 9 and 11 in Ref.~\cite{Octet2}  
(but the effect is less significant for $\Xi^-$).]
Such feature can be interpreted as the enhancement 
of the in-medium charge square radii compared with those in vacuum.
The falloff of the ratio $G_{E}^*/G_{E}$ increases  
when the density increases. 
As for the charge neutral baryons $n$, $\Lambda$, $\Sigma^0$ and $\Xi^0$,
the medium modifications show different features though.
The $\Lambda$ and $\Sigma^0$ systems are 
dominated by the pion cloud contributions 
and the medium modifications are small,
while the $n$ and  $\Xi^0$ systems are dominated by the valence 
quark contributions, and the medium modifications can be significant.
Those medium modifications can be understood by the result 
of the in-medium enhancement of the charge square radii.

\begin{table}[t]
\begin{center}
\begin{tabular}{c  c }
\hline
\hline
   $B$  & $G_{MB}^\ast(0)$ \\
\hline
\hline
 $p$        & 
$ \left[ 1  + \left(\frac{8}{9} \kappa_u   +  \frac{1}{9}  \kappa_d \right)  \right] \frac{M_N}{M_N^\ast}  $ \\
 $n$        & 
$ - \left[ \frac{2}{3}  + \left(\frac{2}{9} \kappa_u   +  \frac{4}{9}  \kappa_d \right)  \right] \frac{M_N}{M_N^\ast} $ 
\\[.35cm]
 $\Lambda$   & 
$ - \frac{1}{3} \frac{M_N}{M_\Lambda^\ast}  - \frac{1}{3} \kappa_s \frac{M_N}{M_N^\ast}$  \\[.35cm]
 $\Sigma^+$  &  $\frac{M_N}{M_\Sigma^\ast} + \left(\frac{8}{9}\kappa_u + \frac{1}{9}\kappa_s  \right) 
\frac{M_N}{M_N^\ast}$ \\
 $\Sigma^0$ & $ \frac{1}{3} \frac{M_N}{M_\Sigma^\ast} + 
\left(\frac{4}{9}\kappa_u - \frac{2}{9}\kappa_d + \frac{1}{9}\kappa_s  \right) 
\frac{M_N}{M_N^\ast}$ \\
 $\Sigma^-$ & $ -\frac{1}{3} \frac{M_N}{M_\Sigma^\ast} - 
\left(\frac{4}{9}\kappa_d - \frac{1}{9}\kappa_s \right) 
\frac{M_N}{M_N^\ast}$ \\[.35cm] 
 $\Xi^0$    &   $ -\frac{2}{3} \frac{M_N}{M_\Xi^\ast} - 
\left(\frac{2}{9}\kappa_u + \frac{4}{9}\kappa_s \right) 
\frac{M_N}{M_N^\ast}$ \\
 $\Xi^-$    &    $ -\frac{1}{3} \frac{M_N}{M_\Xi^\ast} + 
\left(\frac{1}{9}\kappa_d - \frac{4}{9}\kappa_s \right) 
\frac{M_N}{M_N^\ast}$ \\[.2cm]
\hline
\hline
\end{tabular}
\end{center}
\caption{\footnotesize
Explicit expressions for $G_{MB}^\ast(0)$ in terms 
of the octet baryon mass in medium ($M_B^\ast$)
and the free nucleon mass to convert the nucleon natural units.
The quark anomalous magnetic moment values  
are $\kappa_u = 1.711$, $\kappa_d= 1.987$ and 
$\kappa_s = 1.462$~\cite{Octet2}.}
\label{tabGMB0}
\end{table}

Concerning the results of $G_{M}^\ast/G_{M}$, 
they are very similar for all the octet baryons.
[See Figs.~4--11 in Ref.~\cite{Octet2}.]
In vacuum as well as in-medium cases we have falloffs with $Q^2$.
The falloffs are faster for larger nuclear densities.
In the region near $Q^2=0$, however, the magnitude of $G_{M}^\ast/G_{M}$
is, in general, determined by the valence quark 
contributions, which are dominated by the in-medium masses of the baryons.
The expressions for the valence quark 
contributions for the in-medium magnetic form factor $G_M^\ast$
at $Q^2=0$ are presented in Table~\ref{tabGMB0}.
We can notice in Table~\ref{tabGMB0} that $G_{MB}^\ast(0)$
can be decomposed into two main terms:
a term in $\frac{M_N}{M_B^\ast}$
and a term  expressed only in terms of the nucleon masses
($\frac{M_N}{M_N^\ast}$). 
These results are the consequence 
of our extension of the covariant spectator quark model 
to symmetric nuclear matter,
based on a quark current which has the dependence 
of the quark anomalous magnetic moment ($\kappa_q$) 
in the form $\kappa_q/(2 M_N^\ast)$. 
Note that in the case of the nucleon we need 
to take into account only one term.
One has then $\frac{G_M^\ast}{G_M} \propto \frac{M_N}{M_N^\ast}$,
and this means that the $G_M$ is expected to be enhanced in medium.
As for the other baryons 
it is expected that the term in $\frac{M_N}{M_N^\ast}$
has a strong impact on the medium modifications, 
since the factor $\frac{M_N}{M_B^\ast}$ 
is less sensitive to the mass variations in medium.
The magnetic form factors in larger densities 
(or smaller effective masses) are more enhanced 
than those in smaller densities.

\begin{figure*}[t]
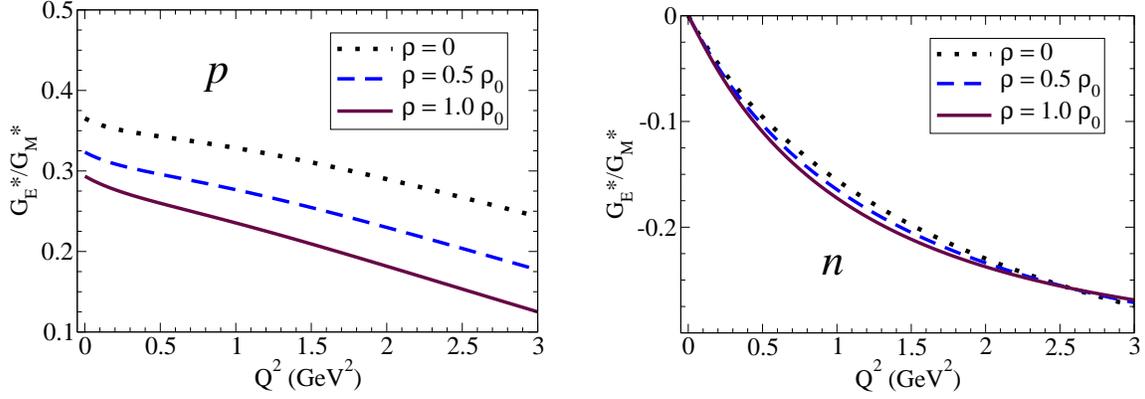

\centerline{
\mbox{
\includegraphics[width=2.8in]{ProtonGEGM.eps} \hspace{.6cm}
\includegraphics[width=2.8in]{NeutronGEGM.eps} }}
\caption{\footnotesize{
EMFF single ratios $G_E^\ast/G_M^\ast$ in symmetric nuclear matter 
for proton (left) and neutron (right), 
where $\rho_0 = 0.15$ fm$^{-3}$~\cite{Octet2}.}}
\label{figNucleonSR}
\end{figure*}

Once we have discussed the properties of the model in medium, 
we can present our estimates for 
the octet baryon $G_{E}/G_M$ double ratios  
for the nuclear matter densities $0.5 \rho_0$ and $\rho_0$ 
with $\rho_0 = 0.15$ fm$^{-3}$.

\section{Octet baryon EMFF double ratios 
{\boldmath $(G^*_E/G^*_M)/(G_E/G_M)$} in symmetric nuclear matter}
\label{secResults}

We present here our estimates of the octet baryon 
electromagnetic form factor double ratios 
$\frac{G^*_E/G^*_M}{G_E/G_M}$ in symmetric nuclear matter.

In the cases of the proton and the neutron, we have experimental data for 
the ratio $G_E/G_M$ in vacuum~\cite{Octet2}.
The estimates for $G_E/G_M$ in vacuum and in symmetric nuclear matter 
for these cases are presented in Fig.~\ref{figNucleonSR}.
The results for the vacuum (dotted line) are in agreement 
with the corresponding experimental data 
(see Ref.~\cite{Octet2}).
The results of the double ratios for the proton and neutron  
can be better understood when we know the results for the proton and 
neutron single ratios $G_E/G_M$ in vacuum and in symmetric nuclear matter.

For the discussion in symmetric nuclear matter, 
it is important to recall that the magnetic form factors are converted into 
the units of nuclear magneton in vacuum, $\hat \mu_N = \frac{e}{2 M_N}$. 

In the case of the nucleon the magnetic 
form factor near $Q^2=0$ can be expressed in the 
form $G_M^\ast \propto 1/M_N^\ast$ and thus the ratio $\frac{G_M^\ast}{G_M}$ 
can be expressed at low $Q^2$ by 
$\frac{G_M^\ast}{G_M^\ast} \approx \frac{M_N}{M_N^\ast}$. 
As for the other baryons, in the case 
when the valence quark contributions are dominant,   
we need to take into account the impact 
of the terms in $\frac{M_N}{M_B^\ast}$ and $\frac{M_N}{M_N^\ast}$
(see Table~\ref{tabGMB0}).
It is worth mentioning that the medium effects 
on $G_{M} (Q^2)$ are not restricted to the factors 
$\frac{M_N}{M_B^\ast}$ and $\frac{M_N}{M_N^\ast}$,
since the form factors 
$F_{1}(Q^2)$ and $F_{2}(Q^2)$ are also modified in the medium. 
The effect of the mass factors is, however,
in general the dominant effect.

For the EMFFs of charge neutral baryons, 
it is also important to notice that, in the low-$Q^2$ region, 
the electric form factor of a baryon $B$ can be expressed 
as $G_{EB} \simeq - \frac{1}{6} r_{EB}^2 Q^2$,
where $r_{EB}^2$ is the electric charge square radius of the baryon $B$.
Then, we can write in the low-$Q^2$ region  
$\frac{G_{EB}^\ast}{G_{EB}} \simeq \frac{r_{EB}^{\ast \, 2}}{r_{EB}^2}$,  
where $r_{EB}^{\ast \, 2}$ is the in-medium electric charge square radius.

\begin{figure*}[t]
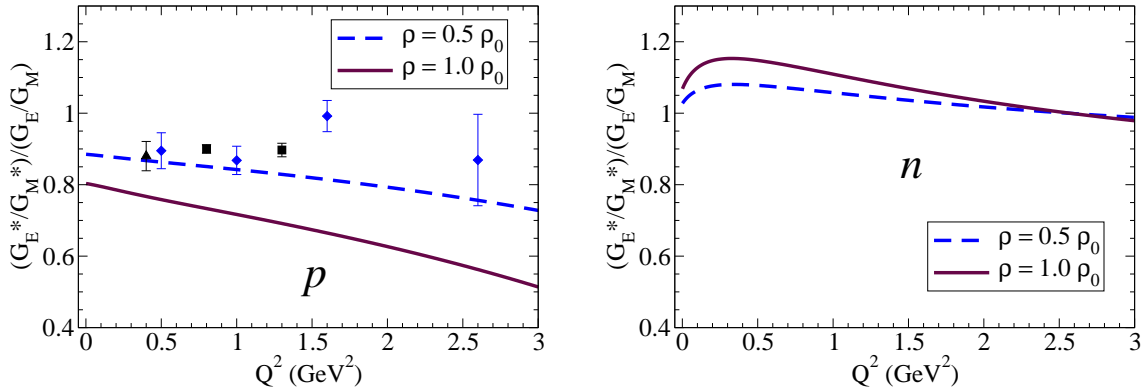

\centerline{\vspace{0.5cm}  }
\centerline{
\mbox{
\includegraphics[width=2.8in]{ProtonDRatio.eps} \hspace{.6cm}
\includegraphics[width=2.8in]{NeutronDRatio.eps} }}
\caption{\footnotesize{
EMFF double ratios in symmetric nuclear matter calculated for the proton (left)
and the neutron (right).
Experimental data for the proton are from Refs.~\cite{Dieterich01,Strauch04,Paolone10}.
}}
\label{figNucleon}
\end{figure*}

Our results for the octet baryon double ratios
$\frac{G_E^\ast/G_M^\ast}{G_E/G_M}$ 
are displayed in Figs.~\ref{figNucleon}--\ref{figLambda-Sigma0}.
We discuss separately each case in the following for  
proton, neutron, $\Sigma^\pm$ and $\Xi^-$, 
and the other charge neutral baryons ($\Lambda, \Sigma^0$ and $\Xi^0$).

\subsection{Proton double ratios}

We start our discussion with the proton   
double ratios displayed in the left panel of Fig.~\ref{figNucleon}.

From the results of the single ratio $G^*_E/G^*_M$, 
the left panel of Fig.~\ref{figNucleonSR}, 
one can see that the single ratio in vacuum ($\rho=0$) decreases 
almost linearly with $Q^2$ (dotted line).
A simple analysis of the result in vacuum for $G_E/G_M$ 
can be made using the low-$Q^2$ expansion discussed in Ref.~\cite{Cloet09}: 
$\frac{G_E}{G_M}  \simeq \sfrac{1}{\mu_p} \left[  
1 - \sfrac{Q^2}{6}(r_{Ep}^{2} -r_{Mp}^{2}) \right]$,  
where $r_{Ep}^{2}$ and $r_{Mp}^{2}$ are, respectively,  
the proton electric charge and magnetic dipole square radii in vacuum. 
This estimate also holds in symmetric nuclear matter 
by replacing the in-vacuum quantities by the corresponding in-medium quantities:  
$\mu_p^\ast$, $r_{Ep}^{\ast \, 2}$ and  $r_{Mp}^{\ast \, 2}$.
The nearly linear falloff with $Q^2$ in vacuum 
is the consequence of the small magnitude of the factor 
$(r_{Ep}^2 -r_{Mp}^2)$, which means that the values 
of $r_{Ep}^2$ and $r_{Mp}^2$ are close. 
Although this analysis is only valid in the low-$Q^2$ region  
(say $Q^2 < 1$ GeV$^2$)~\cite{Cloet09}, it is interesting to notice that 
the approximated linear dependence is observed also for large $Q^2$ 
in the present model.
From Fig.~\ref{figNucleonSR}, one can also conclude that the 
falloff increases with the nuclear matter density $\rho$,
suggesting that $(r_{Ep}^{\ast \, 2} - r_{Mp}^{\ast \, 2})$ 
increases also with the nuclear matter density $\rho$.

Looking now at the double ratios, in the left panel of Fig.~\ref{figNucleon},
we can see that the magnitudes at $Q^2=0$ are less than unity for 
both $0.5 \rho_0$ and $\rho_0$. 
These results are essentially a consequence of the enhancement  
of the in-medium magnetic form factor, 
since $\frac{G_E^\ast}{G_E} \simeq 1$ and 
$\frac{G_M^\ast}{G_M} \approx \frac{M_N}{M_N^\ast}$, near $Q^2=0$.
The $Q^2$ dependence of the double ratios can now be understood 
based on the results for the ratios $G_E/G_M$ and 
the nearly linear falloff with increasing $Q^2$.
The results for the proton double ratios indicate 
that the factor $(r_{Ep}^{\ast \,2} -r_{Mp}^{\ast \, 2})$ is enhanced 
in symmetric nuclear matter.
As a consequence, the falloff of $G_E^\ast/G_M^\ast$ increases in symmetric nuclear matter.

Our results for $0.5 \rho_0$ are similar 
to those in Ref.~\cite{deAraujo18}, for $\rho=0.4 \rho_0$. 
However, our estimates for both $0.5 \rho_0$ and $\rho_0$  
give slower falloff rates with $Q^2$ 
compared with the estimates based on 
the Nambu-Jona-Lasinio model in Ref.~\cite{Cloet09}.
A relativistic light-front constituent quark model~\cite{Chung91} 
predicts an almost constant double ratio.
Other studies predict different $Q^2$ dependence~\cite{Smith04,Chung91,Lu99a}.

\subsection{Neutron double ratios}

We now discuss the results for the neutron double ratios   
presented in the right panel of Fig.~\ref{figNucleon}.
It is worth to recall that, although 
there is no experimental extraction for 
the neutron double ratio at the moment,
the double ratio results are expected from the 
$^2$H$(\vec{e},e^\prime \vec{n})p$  
and $^4$He$(\vec{e},e^\prime \vec{n})^3$He reactions  
in the near future~\cite{Strauch18,Cloet09,PAC35}.

The results for the neutron double ratios are also very interesting.
Contrary to the case of the proton, the neutron double ratios increase 
in the region $Q^2 < 2.8$ GeV$^2$.
The medium effects enhance the neutron double ratios up to 7\%
for $0.5 \rho_0$ and 15\% for $\rho_0$   
near $Q^2 = 0.3$ GeV$^2$, before starting to fall down 
slowly up to $Q^2 \simeq 2.8$ GeV$^2$,  
where the double ratios become closer to the unity.
Beyond the point $Q^2 = 3$ GeV$^2$, the double ratios    
become smaller than unity (not shown in this article). 

\begin{figure*}[t]
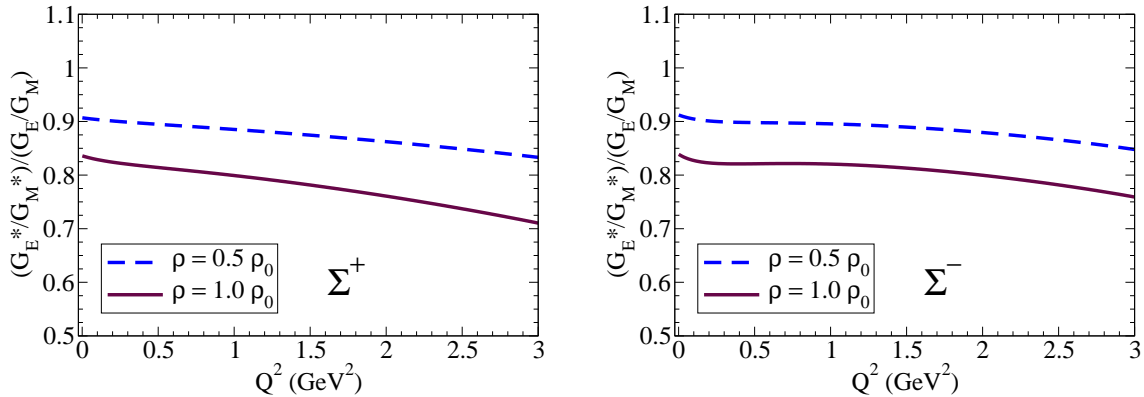

\centerline{\vspace{0.5cm}  }
\centerline{
\mbox{
\includegraphics[width=2.8in]{SigmaP-DRatio.eps} \hspace{.6cm}
\includegraphics[width=2.8in]{SigmaM-DRatio.eps} }}
\caption{\footnotesize{
EMFF double ratios in symmetric nuclear matter calculated for $\Sigma^+$ (left)
and $\Sigma^-$ (right).
}}
\label{figSigma}
\end{figure*}

It is worth noticing that, in general, the amounts of increase 
in the neutron double ratios are smaller than those of the reduction  
in the proton double ratios for the corresponding nuclear densities.
The important conclusion is that, contrary to the proton,
the neutron double ratio $\frac{G_E^\ast/G_M^\ast}{G_E/G_M}$ increases 
in symmetric nuclear matter.
The present result for $\frac{G_E^\ast/G_M^\ast}{G_E/G_M} > 1$  
is consistent with the estimates made in Refs.~\cite{Cloet09,deAraujo18}, 
although the shapes are different. 
The magnitude of the enhancement for  
the neutron double ratios is, in general, smaller than the estimates 
made in other studies with similar nuclear densities~\cite{Cloet09,deAraujo18}. 
 
The significant difference between the double ratios   
of the proton and the neutron for a given density ($0.5 \rho_0$ and $\rho_0$) 
can be understood by the different behaviors of 
$G_E/G_M$ in vacuum as well as in medium.
In the right panel of Fig.~\ref{figNucleonSR}, 
we can see that the neutron single ratio $G_E^\ast/G_M^\ast$ 
decreases monotonically with $Q^2$ starting from  
$Q^2=0$, the point $G_E^\ast/G_M^\ast =0$, for the three densities.
The differences among the different densities are the slopes.
The derivatives increase in absolute values with the density $\rho$. 
The common starting point ($G_E^\ast/G_M^\ast =0$) 
is the consequence of $G_{En} \propto r_{En}^2 Q^2$, near $Q^2=0$.
The slopes of $G_{En}$ and $G_{En}^\ast$
can be estimated by $r_{En}^2$ and $r_{En}^{* \,2}$, respectively.
Those slopes have an important role in the double 
ratios\footnote{The neutron electric charge square radius is negative 
in vacuum and also in symmetric nuclear matter.
Thus, the signs are irrelevant for the discussion,  
since they cancel in $G_E^\ast/G_E$.}.

Since the neutron is a charge neutral baryon,  
the medium modification of $G_E$ 
can be estimated
using the low-$Q^2$ relation:
$\frac{G_{En}^\ast}{G_{En}} \approx \frac{r_{En}^{\ast \, 2}}{r_{En}^2}$.
One can then reduce the analysis of the effects in medium 
to the study of the variation of 
the neutron electric charge square radius in symmetric nuclear matter.
The enhancement (faster falloff) of the electric form factors in medium 
at low $Q^2$ can be understood by the increasing
of the neutron electric charge square radius 
in symmetric nuclear matter.
In this aspect the medium effect in the electric form factor 
of the neutron differs significantly 
from that of the proton ($G_{Ep}^\ast/G_{Ep} \approx 1$).
Looking at the numerical results,
we obtain at the photon point ($Q^2=0$)
$\frac{G_{En}^\ast}{G_{En}} \simeq 1.16$ for $0.5 \rho_0$
and $\frac{G_{En}^\ast}{G_{En}} \simeq 1.33$ for $\rho_0$.
These results show that the electric charge square radius 
increases 16\% for $0.5 \rho_0$, and 33\% for $\rho_0$ 
relative to that in vacuum.
In both cases, we conclude that the 
neutron electric charge square radius is enhanced
in symmetric nuclear matter.
The enhancement increases as the nuclear density increases.

Different from the proton case,
one can conclude that to estimate the medium modifications 
of the neutron EMFF double ratio,
it is not sufficient to take into account 
only the medium modification of the magnetic form factor.
The medium modifications of $G_M$ 
are given by the factor ($\propto \frac{M_N}{M_N^\ast} < 1$).
The impact of the medium modifications of $G_{E}$ are
given by the factor
$\frac{r_{En}^{\ast \, 2}}{r_{En}^2} > 1$,
compensate the effect of $G_M$, and enhance the double ratio
to values larger than unity. 
The combination of the two effects is that 
the neutron double ratio can be approximated 
by $\frac{G_E^\ast/G_M^\ast}{G_E/G_M} \approx 
\frac{r_{En}^{\ast \, 2}}{r_{En}^2} \frac{M_N^\ast}{M_N}$ 
in the low-$Q^2$ region.

The particular $Q^2$ dependence obtained  
for the neutron EMFF double ratio (nonlinear shape)
is a consequence of the properties of the model.
Contrary to the analysis made in Ref.~\cite{Cloet09},
which is restricted to the low-$Q^2$ region, 
our analysis cannot be exclusively restricted to the variables  
$\mu_B^\ast$, $r_{EB}^{\ast  \,2}$ and $r_{MB}^{\ast \,2}$, 
since we take into account the medium modifications in a wider range of $Q^2$.
 
As mentioned in Sec.~\ref{secModel},  
the covariant spectator quark model breaks the $SU(2)$ symmetry  
at the level of the quark EM currents.
The main consequence is that the neutron form factors 
in vacuum are dominated by the valence quark contributions.
In particular the valence quark contribution 
to $G_{En}$ is about 86\% near $Q^2=0$. 
In the extension for the symmetric nuclear matter 
we still have a dominance of the valence quark contributions.
Since the impact of the in-medium effects 
is different for the valence and for the pion cloud components,
we obtain nonlinear results for the neutron EMFF double ratios.
In this aspect the present model differs 
from the usual other models in the literature, 
where the models for nucleons are based on an almost exact $SU(2)$ 
symmetry~\cite{Thomas84,Miller02,Smith04,Chung91}. 


\begin{figure*}[t]
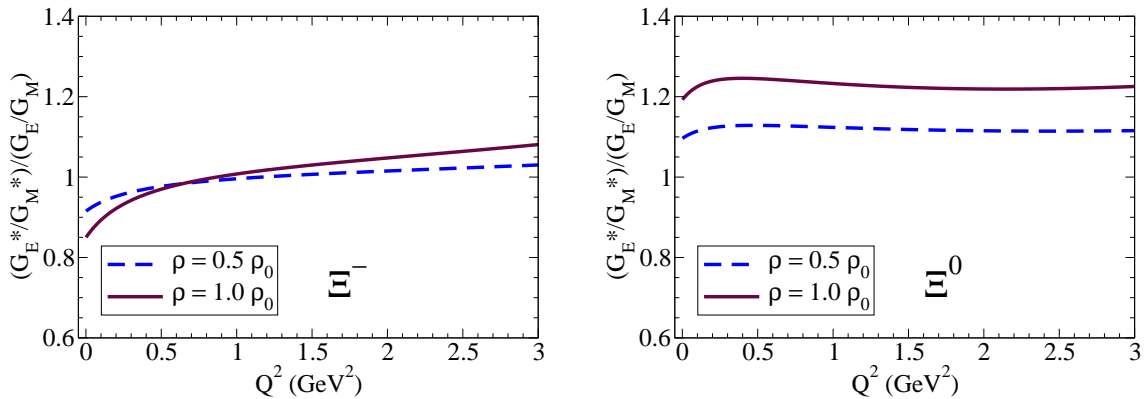

\centerline{\vspace{0.5cm}  }
\centerline{
\mbox{
\includegraphics[width=2.8in]{XiM-DRatio.eps} \hspace{.6cm}
\includegraphics[width=2.8in]{Xi0-DRatio.eps} }}
\caption{\footnotesize{
EMFF double ratios in symmetric nuclear matter calculated for $\Xi^-$ (left)
and $\Xi^0$ (right).
}}
\label{figXi}
\end{figure*}
\begin{figure*}[t]
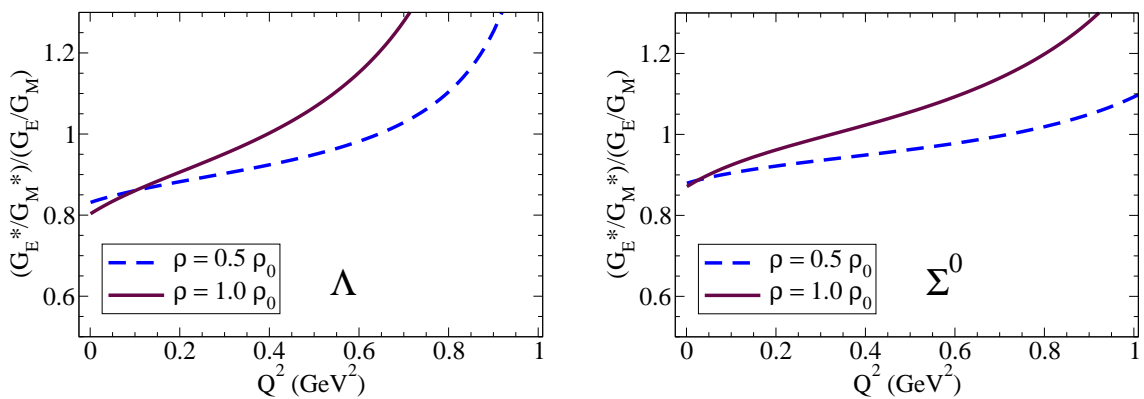

\centerline{\vspace{0.5cm}  }
\centerline{
\mbox{
\includegraphics[width=2.8in]{Lambda-DRatio.eps} \hspace{.6cm}
\includegraphics[width=2.8in]{Sigma0-DRatio.eps} }}
\caption{\footnotesize{
EMFF double ratios in symmetric nuclear matter calculated for $\Lambda$ (left)
and $\Sigma^0$ (right).
}}
\label{figLambda-Sigma0}
\end{figure*}

\subsection{$\Sigma^\pm$ and $\Xi^-$ double ratios}

We now focus on the $\Sigma^\pm$ and  $\Xi^{-}$ 
EMFF double ratios displayed in Figs.~\ref{figSigma} and \ref{figXi}.

The results for the $\Sigma^+$ and $\Sigma^-$ EMFF double ratios  
shown in Fig.~\ref{figSigma} are very similar,  
which may reflect the  $SU(3)$ symmetry.
Our first note goes to the reduction 
in the magnitude of the single ratio $G^*_E/G^*_M$ in symmetric nuclear matter 
compared to that in vacuum. 
The reduction increases with the nuclear density $\rho$. 

The $Q^2$ dependence of the $\Sigma^\pm$ EMFF double ratios 
is very similar to the case of the proton 
except for the slower falloff.
The reduction of the falloff for $\Sigma^\pm$ with $Q^2$,   
in comparison with the proton case,
reflects the presence of strange quarks. 
It is expected that baryons containing more strange quarks 
have smaller medium modifications  
due to the small modifications in the baryon masses.
The small modification in the baryon mass has a small impact 
on the baryon wave function as well as on 
the pion cloud contributions. 
The final effect is the reduction in the falloff of the 
double ratio with $Q^2$. 
As in the proton case, the falloff increases with the nuclear density.

The results of the $\Xi^-$ EMFF double ratios   
are presented in the left panel of Fig.~\ref{figXi}.
One can immediately notice that, contrary to 
the charged baryons discussed so far $p$, $\Sigma^+$ and $\Sigma^-$,
the double ratio increases with $Q^2$, starting from values 
smaller than unity.
The magnitude of the double ratio can be explained  
by observing that $\frac{G_E^\ast}{G_E} \approx 1$ in all ranges of $Q^2$,
as a consequence of the very small contribution from the pion cloud 
contribution for $G_E$.
As for the magnetic form factor which is dominated by 
the valence quark contribution, 
we can consider the approximation 
$\frac{G_M^\ast}{G_M} \approx 0.36 + 0.64 \frac{M_N}{M_N^\ast}$
(see Table~\ref{tabGMB0}),
where the first term includes the smooth  dependence on $M_\Xi^\ast$, 
and the last term, the dependence on $ \frac{M_N}{M_N^\ast}$,
which is enhanced in medium.
For a rough estimate at low $Q^2$, we can use 
$\frac{G_E^\ast/G_M^\ast}{G_E/G_M} \approx  \frac{G_M}{G_M^\ast} \propto \frac{M_N^\ast}{M_N} $.
For larger $Q^2$, one can estimate as $\frac{G_E^\ast/G_M^\ast}{G_E/G_M}\approx 1$, 
reflecting less sensitivity of the strange quarks  
to the medium effects.

\subsection{Charge neutral $\Lambda, \Sigma^0$ and $\Xi^0$ baryon double ratios}

The EMFF double ratio results for the charge neutral baryons 
$\Lambda, \Sigma^0$ and $\Xi^0$
are presented in the right panel of Fig.~\ref{figXi} 
and in Fig.~\ref{figLambda-Sigma0}.

In the $\Lambda$ and $\Sigma^0$ double ratios  
the electric form factors in vacuum $G_E$ vanish at some finite points 
above $Q^2 = 1$ GeV$^2$ (see Figs.~6 and 8 in Ref.~\cite{Octet2}). 
For this reason, there are singularities in  
both double ratios in the region $Q^2 > 1$ GeV$^2$. 
Thus, we restrict to present the results for 
the region $Q^2 \le  1$ GeV$^2$.

Figure~\ref{figLambda-Sigma0} shows that the $Q^2$ dependence of 
the $\Lambda$ and $\Sigma^0$ double ratios are very similar.
The double ratio increases faster in the $\Lambda$ double ratio 
because of the singularity due to the $G_E=0$, 
which happens at lower values of $Q^2$ than that for $\Sigma^0$.

As in the case of the neutron, we need to look at the factor 
$G_E^\ast/G_E$ to understand better the results.
For both the $\Lambda$ and $\Sigma^0$ cases, we have 
$G_E^\ast/G_E \approx 1$, within a 5\% uncertainty for $0.5 \rho_0$
and $\rho_0$, meaning that 
$\frac{G_E^\ast/G_M^\ast}{G_E/G_M} \approx  \frac{G_M}{G_M^\ast}$.
The results in Fig.~\ref{figLambda-Sigma0} reflect then 
the impact of medium modification on $G_M$.
The medium effects are stronger for larger densities.
From the analysis of the valence quark and pion cloud effects 
discussed in Ref.~\cite{Octet2}, we conclude that 
the magnetic form factors are largely dominated 
by the valence quark effects.    
The small medium modifications for  
the electric form factors, $G_E^\ast/G_E \approx 1$, 
on the other hand, are the consequence of the pion cloud dominance.
The pion cloud contributions are only slightly modified in medium.

The results for the $\Xi^0$ EMFF double ratios (right panel of Fig.~\ref{figXi})
show a significant enhancement in comparison 
with that in vacuum, and 
an almost independence of $Q^2$ (nearly constant double ratios).
In symmetric nuclear matter $G_E^\ast/G_M^\ast$ is enhanced  
by about 10\% for $0.5 \rho_0$, and by about 20\% for $\rho_0$.
The nearly constant double ratios,  
for both densities are the consequence of 
an almost cancellation of the $Q^2$ dependence in the division 
of $\frac{G_E^\ast}{G_E}$ by $\frac{G_M^\ast}{G_M}$, 
which means that $\frac{G_E^\ast}{G_E} \propto \frac{G_M^\ast}{G_M}$.

Since the double ratios are nearly constant,  
we can estimate their magnitudes by looking at 
the low-$Q^2$ region, where the $Q^2$ dependence is simplified.
We then obtain $\frac{G_E^\ast}{G_E} \simeq \frac{r_{EB}^{\ast \, 2}}{r_{EB}^2}$, 
and $\frac{G_M^\ast}{G_M} \simeq  -(0.32 + 0.68 \frac{M_N}{M_N^\ast})$,
according to Table~\ref{tabGMB0}.
As in the case of the neutron there is a significant enhancement 
for the electric charge radius in symmetric nuclear matter.
The enhancement is of about 20\% for $0.5 \rho_0$ and 40\% for $\rho_0$.
These effects are partially canceled by the effects of 
the in-medium nucleon mass $\propto \frac{M_N}{M_N^\ast}$ for the two densities, 
which correspond to a 10\% correction for $0.5 \rho_0$ and  20\% for $\rho_0$.
Then the $\Xi^0$ double ratio can be estimated 
in a simple form,  $\frac{G_{E}^\ast/G_{M}^\ast}{G_{E}/G_{M}} 
\propto \frac{r_{EB}^{\ast \, 2}}{r_{EB}^2} \frac{M_N}{M_N^\ast}$.
In the present model, these results are 
the consequence of the dominance of the valence quark contribution,  
since the pion cloud contribution 
is negligible for $\Xi^0$ (see Fig.~10 of Ref.~\cite{Octet2}).

To summarize, our estimates of the $\Xi^0$ EMFF double ratios
are based on the dominance of the valence quark contribution
in both the electric and magnetic form factors.
We then predict a significant enhancement 
of the double ratios in the nuclear medium,  
and a weak $Q^2$ dependence.

\section{Outlook and conclusions}
\label{secConclusions}

We have estimated, for the first time, 
the electromagnetic form factor double ratios 
$\frac{G^*_E/G^*_M}{G_E/G_M}$, 
namely, the in-medium electric to magnetic form factor ratio 
to that ratio in vacuum, for all the octet baryon members.
The estimates are based on the covariant spectator quark model 
plus an $SU(3)$ symmetry-based 
parametrization for the pion cloud contributions.
The extension of the model for symmetric nuclear matter 
is performed using the in-medium inputs for the octet baryons
and  $\rho$- and $\omega$-meson effective masses calculated 
in the quark-meson coupling model,    
as well as the in-medium pion-baryon coupling constants  
obtained using the Goldberger-Treiman relation.
The model parameters in vacuum are calibrated by the
lattice QCD simulation data with large pion masses 
for the valence quark part
and by physical nucleon electromagnetic form factor data
and  octet baryon magnetic moments for the pion cloud part.
Our estimates are covariant and are performed for a wide range of $Q^2$
and not restricted to the low-$Q^2$ region.

One of the main interests of our work was originated from
the different behaviors of the electromagnetic form factor
double ratios for the proton and neutron,
for which, so far, only a few studies have been made.
While the double ratio of the proton decreases with $Q^2$  
as was observed in the JLab experiments, 
that for the neutron is predicted to increase in medium in the low- 
and intermediate-$Q^2$ region ($Q^2 < 2.8$ GeV$^2$).
This prediction can be tested in the near future with the advanced  
polarization-transfer experiment or the developments 
of other new experimental methods.

The first estimates for the neutron electromagnetic form factor 
double ratio has motivated us 
to study the double ratios of the charge neutral 
particles in the octet baryon members.
Our present estimates suggest that, contrary 
to the case of the neutron double ratio,  
those of the $\Lambda$ and $\Sigma^0$ are  
suppressed in medium in the low-$Q^2$ region.
On the other hand, for the large-$Q^2$ region, we expect the  
enhancement of the double ratios for $\Lambda$ and $\Sigma^0$.

For the electromagnetic form factor double ratios of the charged particles 
$\Sigma^-$ and $\Sigma^+$, we predict that they are quenched    
and become less than unity as in the case of the proton.
The $Q^2$ dependence is weaker than the case  
of the proton, mainly because the systems with strange 
quarks are less affected by medium modifications.

As for the $\Xi^-$ and $\Xi^0$ baryons containing two strange quarks,  
we conclude that the electromagnetic form factor double ratios 
have a weaker $Q^2$ dependence compared with 
those of the baryons containing a smaller number of strange quarks,
although we admit that our estimates for the $\Xi^-$ and $\Xi^0$ 
are less precise.

The medium effects as the ones estimated by the present study
have an important impact on the baryons immersed in a nuclear medium,  
in such matter formed in heavy-ion collisions,  
and the one expected to exist in the core of a neutron star and a compact star.
In order to better understand the medium effects,  
the measurement of octet baryon electromagnetic form factor double ratios 
is very important for testing and improving our theoretical estimates.
At the moment, the measurement of the electromagnetic form factor 
double ratios is restricted to the case of the proton
based on the polarization-transfer method,
although the results for the neutron are expected in the near future.
There is then some hope that the polarization-transfer method
or new methods can be used to  directly extract 
the in-medium modifications of octet baryon 
electromagnetic form factors.


\begin{acknowledgments}
G.~R.~was supported by the Funda\c{c}\~ao de Amparo \`a
Pesquisa do Estado de S\~ao Paulo (FAPESP), 
Process No.~2017/02684-5, Grant No.~2017/17020-BCO-JP. 
This work was also partially supported by Conselho Nacional de Desenvolvimento 
Cient\'ifico e Tecnol\'ogico~(CNPq), Brazil, Process No.~401322/2014-9 
(J.~P.~B.~C.~M.), 
No.~308025/2015-6 (J.~P.~B.~C.~M.), No.~308088/2015-8 (K.~T.), 
No.~313063/2018-4 (K.~T.), and No.~426150/2018-0 (K.~T.). 
This work was part of the projects, 
Instituto Nacional de Ci\^{e}ncia e Tecnologia - 
Nuclear Physics and Applications (INCT-FNA), Brazil, Process No.~64898/2014-5, 
and FAPESP Tem\'{a}tico, Process No.~2017/05660-0. 
\end{acknowledgments}

\end{document}